\input phyzzx

\def\cM {{\cal{M}}}

%%%%%%%If you do not have the msbm fonts, delete the following 4 lines
\font\mybb=msbm10 at 12pt
\def\bb#1{\hbox{\mybb#1}}
\def\bZ {\bb{Z}}
\def\bR {\bb{R}}
\def\bE {\bb{E}}

\def\bH {\bb{H}}
\def\hk {hyper-K{\" a}hler}
\def\HK {Hyper-K{\" a}hler}
\def\bfomeg{\omega\kern-7.0pt\omega}
\def\bfOmeg{\Omega\kern-8.0pt\Omega}
\def\bfmu{\mu\kern-8.0pt\mu}
%%%%%%%%%%%%

%%%%%%%%%%%%%%%%%%%%%%%%%%%%%%%%%%%%%%%%%%%%%%%%%%%%%%%%%%%%%%%%%%%%%%%%%%%%%
\REF\pkta{P.K. Townsend, {\sl The eleven-dimensional supermembrane revisited}, 
Phys. Lett. {\bf 350B} (1995) 184.}
\REF\costa{M. Costa and G. Papadopoulos, {\sl Superstring dualities and p-brane
bound states}, hep-th/9612204.}
\REF\BDL{M. Berkooz, M.R. Douglas and R.G. Leigh, {\it Branes Intersecting at
Angles}, Nucl. Phys.  {\bf B480} (1996) 265.}
\REF\BL{V. Balasubramanian and R.G. Leigh, {\sl D-branes, moduli and
supersymmetry}, hep-th/9611165.}
\REF\BB{K. Becker and M. Becker, {\sl M-theory on eight-manifolds},
Nucl. Phys. {\bf B477} (1996) 155.}
\REF\DLPS{M.J. Duff, H. L\" u, C.N. Pope and E. Sezgin, 
{\sl Supermembranes with
fewer supersymmetries}, Phys. Lett. {\bf 371B} (1996) 206.}
\REF\HW{A. Hanany and E. Witten, {\sl Type IIB superstrings, BPS monopoles and
three-dimensional gauge theories}, hep-th/9611230.}
\REF\KL{B.M. Zupnik and D.V. Khetselius, {\sl Three-dimensional extended 
supersymmetry in harmonic superspace}, Yad. Fiz. {\bf 47} (1988) 1147;
H-C. Kao and K. Lee, {\sl Self-dual Chern-Simons Higgs Systems with an
N=3 extended supersymmetry}, Phys. Rev. {\bf D46} (1992) 4691; H-C. Kao, {\sl
Self-dual Yang-Mills Chern-Simons Higgs systems with an N=3 extended
supersymmetry}, Phys. Rev. {\bf D50} (1994) 2881.}
\REF\telaviv{A.A. Tseytlin, {\sl Composite BPS configurations of p-branes in 10
and 11 dimensions}, hep-th/9702163.}
\REF\GKT{J.P. Gauntlett, D.A. Kastor and J. Traschen, {\sl Overlapping Branes
in M-Theory}, Nucl. Phys. {\bf B478} (1996) 544.}
\REF\berg{E. Bergshoeff, M. de Roo, E. Eyras, B. Janssen and J.P. van der
Schaar, {\sl Multiple Intersections of D-branes and M-branes}, hep-th 9612095.}
\REF\PT{G. Papadopoulos and P.K. Townsend, {\sl Intersecting M-branes}, Phys.
Lett. {\bf 380B} (1996) 273.}
\REF\goto{R. Goto, {\sl On \hk\ manifolds of type $A_\infty$}, Geom. Funct.
Anal. {\bf 4} (4) (1994) 424; {\sl On toric \hk\ manifolds}, Kyoto Univ. 
preprint RIMS-818 (1991)}
\REF\PTb{G. Papadopoulos and P.K. Townsend, {\sl Solitons in supersymmetric 
sigma-models with torsion}, Nucl. Phys. {\bf B444} (1995) 245.}
\REF\PP{H. Pederson and Y-S Poon, Commun. Math. Phys. {\bf 117} (1988) 569.} 
\REF\rocek{N. Hitchin, A. Karlhede, U. Lindstr{\" o}m and M. Ro{\v c}ek,
Commun. Math. Phys. {\bf 108} (1987) 537.}

\REF\dancer{R. Bilawski and A.S. Dancer, {\sl The geometry and topology of
toric \hk\ manifolds}, McMaster preprint (1996).}
\REF\LWY{K. Lee, E. Weinberg and P. Yi, {\sl The moduli space of many BPS
monopoles}, hep-th/9602167.}
\REF\manton{G.W. Gibbons and N.S. Manton, {\sl The moduli space metric for
well-separated BPS monopoles}, Phys. Lett. {\bf 356B} (1995) 32.} 
\REF\gibrub{G.W. Gibbons and P.J. Ruback, Commun. Math. Phys. {\bf 115} (1988)
267.}
\REF\wang{M.Y. Wang, {\sl Parallel spinors and parallel forms}, Ann. Global.
Anal. Geom. {\bf 7} (1989) 59.}

\REF\pktb{P.K. Townsend, {\sl Four lectures on M-theory}, hep-th/9612121.}
\REF\khuri{R.R. Khuri, {\sl  Remark on string solitons},
Phys. Rev. {\bf D48} (1993) 2947; N.D. Lambert, {\sl Heterotic
p-branes from massive sigma models}, Nucl. Phys. {\bf 477} (1996) 141.}
\REF\dgt{M.J. Duff, G.W. Gibbons and P.K. Townsend, {\sl Macroscopic 
superstrings as interpolating solitons}, Phys. Lett. {\bf 332 B} (1994) 321.}
\REF\ght{G.W. Gibbons, G.T. Horowitz and P.K. Townsend, 
{\sl Higher-dimensional 
resolution of dilatonic black hole singularities}, 
Class. Quantum Grav. {\bf 12}		
(1995) 297.}
\REF\DNP{M.J. Duff, B.E.W. Nilsson and C.N. Pope, {\sl Kaluza-Klein
Supergravity}, Phys. Rep. {\bf 130} (1986) 1.}
\REF\lpt{H. L{\" u}, C.N. Pope and P.K. Townsend, {\sl Domain walls from 
anti-de
Sitter space}, Phys. Lett. {\bf 391B} (1997) 39.}

\REF\BHO{E. Bergshoeff, C.M. Hull and T. Ort\'\i n, {\sl Duality in the type II
superstring effective action}, Nucl. Phys. {\bf B451} (1995) 547.}
\REF\tseytlin{A. Tseytlin, {\sl Harmonic superpositions of M-branes}, Nucl.
Phys. {\bf B475} (1996) 149.}
\REF\gibbons{G.W. Gibbons, P. Rychenkova and R. Goto, {\sl \HK\ quotient
construction of BPS monopole moduli spaces}, hep-th/9608085.}
\REF\ruback{G.W. Gibbons and P.J. Ruback, {\sl Winding strings, Kaluza-Klein
monopoles and Runge-Lenz vectors}, Phys. Lett. {\bf 215B} (1988) 653.}
\REF\tseytwo{A.A. Tseytlin, {\sl No force condition and BPS combinations of
p-branes in 11 and 10 dimensions}, hep-th/9609212.}

%%%%%%%%%%%%%%%%%%%%%%%%%%%%%%%%%%%%%%%%%%%%%%%%%%%%%%%%%%%%%%%%%%%%

\Pubnum{ \vbox{ \hbox{DAMTP-R/97/10}\hbox{QMW-PH-97-9}\hbox{NI-970015}
\hbox{hep-th/9702202}} }
\pubtype{}
\date{}

\titlepage

\title {\bf \HK\ Manifolds and Multiply Intersecting Branes}

\author{J.P. Gauntlett\foot{Current address: The Isaac Newton
Institute, 20 Clarkson Rd., CB3 0EH, Cambridge, U.K.}}
\address{Dept. of Physics, Queen Mary and Westfield College,  
\break
Mile End Rd., London E1 4NS, U.K.}
\andauthor{G.W. Gibbons ${}^\star$, G. Papadopoulos and P.K. Townsend
${}^\star$}
\address{DAMTP, University of Cambridge,
\break
Silver St., Cambridge CB3 9EW, U.K.}

\abstract{Generalized membrane solutions of D=11 supergravity, for which
the transverse space is a toric hyper-K{\" a}hler manifold, are shown to have 
IIB duals representing the intersection of parallel 3-branes with 5-branes
whose orientations are determined by their $Sl(2;\bZ)$ charge vectors. These IIB
solutions, which generically preserve 3/16 of the supersymmetry, can be further
mapped to solutions of D=11 supergravity representing the intersection of
parallel membranes with any number of fivebranes at arbitrary angles.
Alternatively, a subclass (corresponding to non-singular D=11 solutions) can be
mapped to solutions representing the intersection on a string of any number of
D-5-branes at arbitrary angles, again preserving 3/16 supersymmetry, as we
verify in a special case by a quaternionic extension of the analysis of
Berkooz, Douglas and Leigh. We also use similar methods to find new 1/8
supersymmetric solutions of orthogonally intersecting branes.}

\endpage
%\pagenumber=1

\chapter{Introduction}

A number of recent developments have underscored the importance of
supersymmetric intersecting p-brane configurations of M-theory and superstring
theories for a variety of phenomena of physical interest. Much of the work on
this subject has been concerned with the realization of intersecting p-brane
configurations as solutions of the relevant effective supergravity theory.
The solutions so far considered have been restricted to those
representing orthogonal intersections. Furthermore, most are `delocalized' in
some directions, with the consequence that their interpretation as intersecting
branes is not completely straightforward. In this paper we report on some
progress towards lifting these restrictions. A novelty of many of the new
solutions we find is that they preserve 3/16 supersymmetry, a
fraction not obtainable from orthogonal intersections. Our method derives
from considering certain solutions of D=11 supergravity so it has the additional
advantage that an M-theory setting is automatically provided. Moreover, many
of these D=11 supergravity solutions are non-singular, thereby providing
examples of how the singularities of at least some intersecting brane solutions
of type II superstring theory are resolved in M-theory.

Using M-theory and superstring dualities, any intersecting brane solution can be
obtained from some solution of D=11 supergravity, although it does not follow
from this fact that the D=11 solution will also have an intersecting brane
interpretation. Conversely, solutions of D=11 supergravity with some other
interpretation, or with no obvious interpretation at all, may be interpretable
as intersecting brane solutions after reduction to D=10 and possible
dualization. If one regards a single brane as a degenerate case of an
intersecting brane configuration then a case in point is the solution of D=11
supergravity for which the 11-dimensional metric is the product of 7-dimensional
Minkowski space with the (\hk) Euclidean Taub-Nut metric. Since  the latter
metric admits a U(1) isometry and is asymptotically flat, this solution can be
reduced to D=10 where it becomes the D-6-brane solution of IIA supergravity
[\pkta]. Similarly, the analogous Euclidean Taub-Nut solution of D=10 IIA(B)
supergravity is T-dual to an NS-5-brane solution of IIB(A) supergravity
[\costa]. 

This raises an obvious
question: does the product of two Euclidean Taub-Nut spaces, which is an
8-dimensional asymptotically flat \hk\ manifold with holonomy $Sp(1)\times
Sp(1)$, have an analogous interpretation when used as a solution of D=11
supergravity? The same question can also be asked of any 8-dimensional
asymptotically flat \hk\ manifold, for which the holonomy is generically
$Sp(2)$. One purpose of this paper is to provide an answer to this question for
asymptotically flat 8-dimensional `toric' \hk\ manifolds, i.e. those
admitting a triholomorphic $T^2$ isometry, by showing that the associated
solutions of D=11 supergravity are mapped to intersecting 5-brane solutions of
IIB supergravity. The generic IIB solution found this way preserves 3/16
of the supersymmetry of the IIB vacuum solution and is naturally interpreted as
an arbitrary number of 5-branes with pairwise intersections at angles
determined by their $Sl(2;\bZ)$ charges. 

The fact that these solutions generically preserve 3/16 supersymmetry derives in
the first instance from the fact that \hk\ 8-manifolds generically have holonomy
$Sp(2)$, because this implies the preservation of 3/16 supersymmetry by the D=11
supergravity solution; the triholomorphicity of the $T^2$ isometry then ensures
that this feature is maintained under dimensional reduction to D=10 and
subsequent T-duality. In the special case in which the 8-metric is the metric
product of two 4-metrics the holonomy is reduced to $Sp(1)\times Sp(1)$ and the
corresponding solutions of D=11 supergravity, which preserve 1/4 of the
supersymmetry, are mapped under duality to solutions representing any number of
parallel D-5-branes orthogonally intersecting, or overlapping, any number of
parallel NS-5-branes on a 2-brane. 

These IIB solutions can now be mapped back to D=11 to give new intersecting
M-5-brane solutions in which an arbitrary number of
M-5-branes intersect at arbitrary angles while still preserving 3/16
supersymmetry. Alternatively, a series of duality transformations leads to a
class of solutions of IIB supergravity representing the intersection on a
string of an arbitrary number of D-5-branes, again at arbitrary angles and
preserving 3/16 of the supersymmetry. A potentially useful feature of these
solutions is that, since they involve only D-branes, it should be possible to
further analyse them using string perturbation theory. A special case
represents just two D-5-branes intersecting on a string with one rotated by an
arbitrary angle relative to the other. When the D-5-branes are orthogonal they
preserve 1/4 of the supersymmetry, as expected, so we learn from the more
general solution that a rotation away from orthogonality may be such as to
preserve 3/4 of the original supersymmetry. One might have expected that any
deviation from orthogonality would break all supersymmetries, but it has been
shown previously by other methods that this is not necessarily the case
[\BDL,\BL]. We adapt these methods to our case to verify that the fraction of
supersymmetry preserved, relative to the vacuum, is 3/16. 

The starting point for all the above results are non-singular solutions of D=11
supergravity for which the D=11 4-form field strength is zero and the 11-metric
takes the form
$$
ds^2 = ds^2(\bE^{2,1}) + ds^2_8
\eqn\introa
$$
where $\bE^{2,1}$ is D=3 Minkowski space and $ds^2_8$ is a complete toric \hk\
8-metric. This solution is actually a special case of a `generalized membrane'
solution for which
$$
\eqalign{
ds^2 &= H^{-{2\over3}}ds^2(\bE^{2,1}) + H^{1\over3}ds^2_8\cr
F &= \pm\omega_3 \wedge dH^{-1} }
\eqn\introb
$$
where $\omega_3$ is the volume form on $\bE^{2,1}$ and $H$ is a 
$T^2$-invariant\foot{This condition on $H$ is
needed for our applications; it is not needed to solve the D=11 supergravity
equations. Solutions of the form of \introb\ have been found previously in the
context of KK theory (see [\BB] for an M-theory update). Generalized
membrane solutions of a rather different type, but preserving 3/16
supersymmetry, have also been found [\DLPS].} harmonic function on the \hk\
8-manifold. Provided the sign of the expression for the 4-form $F$ in \introb\
is chosen appropriately, the solution with $F\ne0$ breaks no more
supersymmetries than the solution \introa\ with $F=0$. 

Since the `generalized membrane' solution \introb\ of D=11 supergravity admits
the action of a torus we can convert it to a solution of IIB D=10 supergravity,
as in the pure \hk\ case. The resulting IIB solution generalizes the previous
one in that the 2-brane overlap of the 5-branes now has the interpretation as
the intersection (or, possibly, the boundary) of a D-3-brane with the
5-branes. In the case of orthogonal intersections these are just the
configurations used by Hanany and Witten (HW) in their study of D=3
supersymmetric gauge theories [\HW]. Actually, our supergravity solutions do
not quite correspond to the HW configurations because they are translationally
invariant along the direction in the 3-brane connecting the 5-branes. In
another respect, however, our solutions are more general because they include
configurations of non-orthogonal 5-branes preserving 3/16 supersymmetry,
leading to N=3, rather than N=4, supersymmetry on the D=3 intersection.  These
configurations are therefore of possible relevance to the dynamics of D=3
gauge theories with N=3 supersymmetry [\KL]. 

Given a solution representing a 3-brane intersecting IIB 5-branes, we can
T-dualize along a direction in the 2-brane intersection to arrive at a new IIA
configuration which can then be lifted back to D=11. The resulting solution of
D=11 supergravity, which generically preserves 3/16 supersymmetry, can be
interpreted as an M-theory membrane intersecting, on a string, any number of
M-5-branes at arbitrary angles. The special case in which just two M-5-branes
intersect orthogonally is itself a new solution\foot{This solution has been 
found independently by Tseytlin [\telaviv].}, preserving 1/4 of the vacuum
supersymmetry, that generalizes the `two M-5-branes overlapping on a string'
solution of [\GKT]. 

As a further example of how \hk\ manifolds lead via duality to new intersecting
branes we consider a `generalized overlapping fivebrane' solution of D=11
supergravity of the form
$$
\eqalign{
ds^2_{11} &= (H_1H_2)^{2\over3}\bigg[(H_1H_2)^{-1} ds^2(\bE^{1,1}) +
H_1^{-1}ds^2_4 (M_2) +  H_2^{-1} ds^2_4 (M_1) + 
dz^2\bigg] \cr
F&= \big[*_1 dH_1 + *_2 dH_2\big] \wedge dz }
\eqn\introc
$$
where $H_1$ and $H_2$ are harmonic functions on the respective
asymptotically-flat \hk\ 4-manifolds $M_1$ and $M_2$, each with $U(1)$
isometry, and $*_i$ indicates the Hodge dual on $M_i$. The `two M-5-branes
overlapping on a string' solution is now recovered as the special case for which
$M_1$ and $M_2$ are both chosen to be
$\bE^4$. If one or both $M_1$ and $M_2$ are taken to be Euclidean Taub-Nut (for
example) then 1/8 of the supersymmetry is preserved for appropriate relative
signs of the 5-brane charges. Since this `generalized overlapping fivebrane'
solution still has  a triholomorphic
$T^2$ isometry it can be mapped to a solution of IIB supergravity preserving the
same fraction of supersymmetry. We show that it maps to a configuration of
orthogonally intersecting D-5-branes and NS-5-branes. It can then be mapped back
to D=11 to yield a new\foot{For example, it is not included in a recent
classification of orthogonally intersecting brane solutions [\berg].} 1/8
supersymmetric solution of D=11 supergravity representing (in the simplest
case) four orthogonally intersecting M-5-branes. Remarkably, this can be
further generalized to include a pair of intersecting membranes, still
preserving 1/8 supersymmetry. 

We have been using the terms `intersecting' and `overlapping' interchangeably
in the above discussion, but there is of course a distinction between them.
The possibility of an `overlapping' brane interpretation arises whenever the
branes are potentially separable in one or more directions. If two branes
intersect one expects the intersection to appear as a physical intersection in
the worldvolume field theory of each brane; this leads, for instance, to the
`$(p-2)$-brane intersection' rule [\PT]. The solutions considered here are
typically translationally invariant in the one direction that potentially
separates different branes, so the issue of whether the branes are actually
intersecting or merely overlapping is left unresolved. However, the fact that
the `overlapping' M-theory 5-brane solution of [\GKT] has a generalization in
which the common 1-brane is naturally interpreted as the intersection of each
5-brane with a membrane makes it also natural to suppose that one is
left with a mere overlap when the membrane is removed. In any case, we shall
find it convenient to adopt this point of view here in order to avoid
confusion between different types of solution. For example, we shall refer to
HW-type configurations of IIB D-5-branes and NS-5-branes without 3-branes as
`overlapping' branes whereas we shall refer to the more general configuration
{\it with} 3-branes as `intersecting' branes. 

\chapter{Toric \HK\ manifolds}

We are principally interested here in 8-dimensional \hk\ manifolds with a
tri-holomorphic $T^2$ isometry, but we shall consider these as special cases of
$4n$-dimensional \hk\ manifolds $\cM_n$ with a tri-holomorphic $T^n$ isometry.
We shall refer to them as toric \hk\ manifolds [\goto]. Let 
$$
X^i ={\partial\over\partial\varphi_i} \qquad (i=1,\dots,n)
\eqn\onea
$$
be the $n$ commuting Killing vector fields. They are tri-holomorphic if the
triplet of K{\" a}hler 2-forms $\bfOmeg$ is $\varphi_i$-independent,
i.e. if
$$
{\cal L}_i \bfOmeg =0\ ,
\eqn\oneb
$$
where ${\cal L}_i$ is the Lie derivative with respect to $X^i$. The general 
toric \hk\  $4n$-metric has the local form
$$
ds^2 = U_{ij}\, d{\bf x}^i\cdot d{\bf x}^j + 
U^{ij}(d\varphi_i + A_i)(d\varphi_j + A_j)
\eqn\onec
$$
where $U_{ij}$ are the entries of a positive definite symmetric $n\times n$
matrix function $U$ of the $n$ sets of coordinates ${\bf x}^i = \{x_r^i\, ;
r=1,2,3\}$ on each of $n$ copies of $\bE^3$, and $U^{ij}$ are the entries of
$U^{-1}$. The $n$ 1-forms $A_i$ have the form
$$
A_i = d{\bf x}^j\cdot \bfomeg_{ji}
\eqn\oned
$$
where $\bfomeg$ is a triplet of $n\times n$ matrix functions\foot{Of no
particular symmetry; it was incorrectly stated in [\PTb] that these matrix
functions are symmetric.} of the $n$ sets of $\bE^3$ coordinates. The three
K{\" a}hler 2-forms are [\PTb] 
$$
\bfOmeg = (d\varphi_i + A_i) d{\bf x}^i - {1\over 2} U_{ij}\, d{\bf x}^i \times
d{\bf x}^j
\eqn\oneda
$$
where $\times$ denotes the standard vector product in $\bE^3$, the exterior
product of forms being understood (e.g. the 3-component of $d{\bf x}^i\times
d{\bf x}^j$ is $2dx_1^i\wedge dx_2^j$). 

For some purposes it is convenient to introduce a (non-coordinate) frame in 
which
the components of both the metric and $\bfOmeg$, and hence the complex
structures, are constant. To do so we write $U$ as
$$
U=K^T K
\eqn\trihola
$$
for some non-singular matrix $K$ (which is not unique because it may be
multiplied on the left by an arbitrary $SO(n)$ matrix). We then define $3n$ 
legs of a $4n$-bein by
$$
{\bf E}_i = K_{ij} d{\bf x}^j
\eqn\triholb
$$
and the remaining $n$ legs by
$$
E^i= (d\phi_j+ A_j) K^{ji}
\eqn\triholc
$$
where $K^{ij}$ is the inverse of $K_{ij}$. This $4n$-bein defines a new frame
in which the metric is
$$
ds^2 = \delta^{ij}\, E_iE_j + {1\over2}\delta_{ij}\, {\bf E}^i\cdot {\bf E}^j
\eqn\metrica
$$
and the triplet of K{\" a}hler 2-forms is
$$
\bfOmeg= 2E^i{\bf E}_i - {\bf E}_i \times {\bf E}_j\ .
\eqn\trihold  
$$
In this frame the triplet of complex structures ${\bf J}$ are simply a set of
three constant $4n\times 4n$ matrices

The conditions on $U_{ij}$ and $A_i$ required for the metric to be \hk, and for
the closure of $\bfOmeg$, are most simply expresed as the constraint [\PP]
$$
F_{jk}^{rs}{}_i = \varepsilon^{rst} \partial^t_jU_{ki} 
\eqn\onee
$$
on the 2-form `field strengths' $F_i=dA_i$, for which the components are
$$
F_{jk}^{rs}{}_i = \partial^r_j\omega^s_{ki} - \partial^s_k\omega^r_{ji}\ ,
\eqn\onef
$$
where we have introduced the notation
$$
{\partial\over \partial x_r^i} =\partial^r_i\ .
\eqn\oneg
$$
The constraint \onee\ implies that 
$$
\partial^t_{[j}U_{k]i} =0\ ,
\eqn\oneh
$$
while the `Bianchi' identity $dF_i\equiv 0$ implies that the matrix $U$
satisfies [\rocek]
$$
{\bf \partial}_i\cdot {\bf \partial}_j\, U =0\qquad (i,j=1,\dots,n)\ .
\eqn\onei
$$
Given a solution of these equations, the 1-forms $A_i$ are determined up to a
gauge transformation of the form $A_i\rightarrow A_i + d\alpha_i(x)$, which can
be compensated by a change of $\varphi_i$ coordinates. Thus, the determination
of \hk\ metrics of the assumed type reduces to finding solutions  of \oneh\ and
\onei. We shall refer to these equations as the `\hk\ conditions'.

Note that \onei\ implies that
$$
U^{ij}\, \partial_i\cdot\partial_j\, U =0\ ,
\eqn\onej
$$
which is equivalent to the statement that each entry of $U$ is a harmonic
function on ${\cal M}_n$. To see that this is so, we observe that the Laplacian,
when restricted to acting on $T^n$-invariant functions, is
$$
\eqalign{
\nabla^2 &=(\det U)^{-1}\partial_i \cdot(\det U U^{ij}\partial_j)\cr
&= U^{ij}\, \partial_i\cdot\partial_j \ ,}
\eqn\onek
$$
where the second line follows from \oneh. Since $U$ is $T^n$-invariant, it
follows that \onej\ is equivalent to $\nabla^2 U=0$. Of course, this is far
from being a complete characterization of $U$. 

One obvious solution of the \hk\ conditions, which may be considered to
represent the `vacuum', is constant $U$. We shall denote this constant
`vacuum matrix' by $U^{(\infty)}$ (we shall see in due course that this
terminology is appropriate for the applications we have in mind). The
corresponding `vacuum metric' is
$$
ds^2 = U^{(\infty)}_{ij}d{\bf x}^i\cdot d{\bf x}^j +
U_{(\infty)}^{ij}d\varphi_i d\varphi_j\ .
\eqn\onela
$$
For our applications we shall insist that $\varphi_i$ be periodically
identified with the standard identification 
$$
\varphi_i \sim \varphi_i + 2\pi \qquad (i=1,\dots,n)\ .
\eqn\onelb
$$
Thus, the `vacuum manifold' is $\bE^{3n}\times T^n$. We shall wish to consider
only those \hk\ manifolds that are asymptotic to $\bE^{3n}\times T^n$, with the
above metric and identifications, as $|{\bf x}^i|\rightarrow \infty$ for all
$i$. Thus, the moduli space of `vacua' may be identified with the set of flat
metrics on $T^n$. This in turn may be identified with the double coset space
$$
Sl(n;\bZ)\backslash Gl(n;\bR)/SO(n)\ .
\eqn\onelc
$$

Non-vacuum \hk\ metrics in this class can be found by superposing the constant
solution $U^{(\infty)}$ of the \hk\ conditions with some linear combination of
solutions of the form
$$
U_{ij}[\{ p\},{\bf a}] = {p_ip_j\over 2|\sum_k \, p_k{\bf x}^k\, - {\bf
a}|}
\eqn\onelc
$$ 
where $\{p_i,\, i=1,\dots,n\}$ is a set of $n$ real numbers and ${\bf a}$ is
an arbitrary 3-vector. Any solution of this form may be associated with a
$3(n-1)$-plane in $\bE^{3n}$, specified by the 3-vector equation
$$
\sum_{k=1}^n p_k {\bf x}^k = {\bf a}\ .
\eqn\oneld
$$
The associated \hk\ $4n$-metric is non-singular provided that the parameters
$\{p\}$ are coprime integers. We shall henceforth assume that $\{p\}$ {\it
denotes an ordered set of coprime integers} and we shall refer to this set as a
`p-vector'. The general non-singular metric may now be found by linear
superposition. For a given p-vector we may superpose any finite number
$N(\{p\})$ of solutions with various distinct 3-vectors $\{{\bf a}_m(\{p\});\
m=1,\dots,N\}$. We may then superpose any finite number of such solutions.
This construction yields a solution of the \hk\ conditions of the form
$$
U_{ij} = U^{(\infty)}_{ij} + \sum_{\{p\}}\sum_{m=1}^{N(\{p\})}
U_{ij}[\{ p\},{\bf a}_m(\{p\})]\ .
\eqn\onele
$$
Since each term in the sum is associated to a $3(n-1)$-plane in $\bE^{3n}$, any
given solution is specified by the angles and distances between some finite
number of mutually intersecting $3(n-1)$-planes [\dancer]. It can be shown that
the resulting \hk\ $4n$-metric is complete provided that no two intersection
points, and no two planes, coincide. We demonstrate this in an appendix by
means of the \hk\ quotient construction [\rocek]. 

A feature of this class of \hk\ $4n$-metrics is that it is $Sl(n;\bZ)$
invariant, in the sense that, given $S\in Sl(n;\bZ)$, the $Sl(n;\bZ)$
transformation
$U\rightarrow S^T US$ takes any solution of the \hk\ conditions of the form
\onele\ into another one of this form. This would be true for $S\in Sl(n;\bR)$
if the allowed p-vectors were arbitrary, but the restrictions imposed on them by
completeness of the metric restricts $S$ to lie in the discrete $Sl(n;\bZ)$
subgroup. To see this we note that a $3(n-1)$-plane defined by the p-vector
$\{p\}$ is transformed into one defined by the new p-vector $S\{p\}$ whose
components are again coprime integers only if $S\in Sl(n;\bZ)$.

It will prove convenient to have some simple examples of complete toric \hk\
$4n$-metrics. The simplest examples are found by supposing $\Delta U \equiv
U-U^{(\infty)}$ to be diagonal. For example,
$$
U_{ij} = U^{(\infty)}_{ij} +  \delta_{ij} \, {1\over 2|{\bf x}^i|}\ .
\eqn\onem
$$
\HK\ metrics with $U$ of this form were found previously on the moduli space of
$n$ distinct fundamental BPS monopoles in maximally-broken rank $(n+1)$ gauge
theories [\LWY] (see also [\manton]). For this reason we shall refer to them
as  `LWY metrics'. For the special case in which not only $\Delta U$ but also
$U^{(\infty)}$ is diagonal then $U$ is diagonal and the LWY metrics reduce to
the metric product of $n$ Euclidean Taub-Nut metrics. There is also a
straightforward `multi-centre' generalization of the LWY metrics which reduces
when $U^{(\infty)}$ is diagonal to the metric product  of $n$ cyclic ALF spaces
(see e.g. [\gibrub]). Whenever $\Delta U$ is diagonal we may choose $\bfomeg$ of
\oned\ such that $A_i$ is a 1-form on the $i$th Euclidean 3-space satisfying
$$
F_i = \star dU_{ii}\qquad (i=1,\dots,n)
\eqn\onen
$$
where $\star$ is the Hodge dual on $\bE^3$. 

For our applications we shall also need to know something about covariantly
constant spinors on \hk\ manifolds. We first note that if the holonomy of a
$4n$-dimensional \hk\ manifold is strictly $Sp(n)$ (rather than a proper
subgroup of it) then there there exist precisely $(n+1)$ covariantly constant
$SO(4n)$ spinors [\wang]. As we shall explain in more detail in the following
section for $n=2$, these spinors arise as singlets in the decomposition of the
spinor representation of $SO(4n)$ into representations of $Sp(n)$. It will be 
important in our applications for these covariantly constant spinors to be
independent of the $T^n$ coordinates. Fortunately this is a consequence of the
triholomorphicity of the $T^n$ Killing vector fields. This can be seen as
follows. Because X is Killing, its covariant derivative $\nabla X$ is an
antisymmetric $4n\times 4n$ matrix, i.e. it takes values in the Lie algebra
$so(4n)$. Let $\Psi$ be a field transforming according to a representation $R$
of $SO(4n)$ and let $R(\nabla X)$ be the representative of $\nabla X$ in the
corresponding representation of $so(4n)$. The Lie derivative of $\Psi$ along
$X$ is then 
$$
{\cal L}_X \Psi = i_X \nabla \Psi + R(\nabla X)\Psi\ .
\eqn\triholf
$$
For a covariantly constant spinor $\eta$ we therefore have that
$$
{\cal L}_X \eta = {1\over 4}\Gamma^{ab} (\nabla X)_{ab}\, \eta\ .
\eqn\triholg
$$
The condition that $X$ be triholomorphic, when combined with the covariant
constancy of the complex structures ${\bf J}$, implies that
$$
[\nabla X, {\bf J}] =0
\eqn\trihole
$$ 
This implies that $\nabla X$ actually takes values in the $sp(n)$ subalgebra
of $so(4n)$, but covariantly constant spinors are $Sp(n)$ singlets, so the right
hand side of \triholg\ vanishes and we deduce that ${\cal L}_X \eta=0$, as
claimed.

Only the $n=2$ cases of the above class of \hk\ manifolds will be needed in our
applications. Moreover, for these applications we may restrict $U^{(\infty)}$
to be such that
$$
\det U^{(\infty)} =1\ ,
\eqn\newa
$$
so that the moduli space of `vacua' is 
$$
Sl(2;\bZ)\backslash Sl(2;\bR)/SO(2)\ .
\eqn\newb
$$
Let us first consider the case in which the metric is determined
by just two intersecting 3-planes with p-vectors $\{p\}$ and $\{p'\}$. We can
choose the intersection point to be at the origin of $\bE^6$, so $U$ is given by
$U=U^{(\infty)} + \Delta U$, where
$$
2\Delta U_{ij}=  {p_ip_j\over |p_1{\bf x}^1 + p_2{\bf x}^2|}
+ {p'_ip'_j\over |p'_1{\bf x}^1 + p'_2{\bf x}^2|}\ .
\eqn\newc
$$
The orientation of two 3-planes in $\bE^6$ is specified by the $3\times 3$
matrix 
$$
M^{rs}= (U^{(\infty)})^{ij}\, {\bf n}_i^{(r)} \cdot {\bf m}_j^{(s)}\ ,
\eqn\newd
$$
where $\{{\bf n}_i^{(r)};\, r=1,2,3\}$ are three linearly independent
unit normals to one 3-plane and $\{{\bf m}_i^{(s)};\, s=1,2,3\}$ are three
linearly independent unit normals to the other one. The choice of each set of
three unit normals is irrelevant, so we are free to choose them in such a
way that $M$ is diagonal. Thus, the relative configuration of the two 3-planes
is specified, in principle, by three angles. In our case, however, $M$ is
$SO(3)$ invariant as a consequence of the $SO(3)$ invariance of the conditions
specifying each 3-plane, so 
$$
M^{rs}= (\cos\theta) \delta^{rs}
\eqn\newe
$$
where
$$
\cos\theta = {p\cdot p'\over \sqrt{p^2 p'^2}}\ ,
\eqn\onepa
$$
with inner product
$$
p\cdot q = (U^{(\infty)})^{ij}p_iq_j\ .
\eqn\onepb
$$
When $p=(1,0)$ and $q=(0,1)$, as is the case for the LWY metrics, \onepa\
reduces to
$$
\cos\theta = -{U^{(\infty)}_{12}\over
\sqrt{U^{(\infty)}_{11}U^{(\infty)}_{22}}}\ .
\eqn\onepc
$$

Observe that the formula \onepa\ for the angle between two 3-planes
is $Sl(2;\bR)$-invariant, so given any particular two-plane solution we could
always choose to evaluate the angle between them by making an $Sl(2;\bR)$
transformation of coordinates to bring $U^{\infty}$ to the identity matrix. In
such coordinates $U$ is diagonal, and the metric is therefore the direct product
of two $4$-metrics, whenever the two 3-planes are orthogonal. Thus,
orthogonality of the two 3-planes implies a reduction of the holonomy from
$Sp(2)$ to
$Sp(1)\times Sp(1)$. Non-orthogonality of the two 3-planes does not so
obviously imply that the holonomy is $Sp(2)$ but we have verified, by 
computation of the curvature tensor, that the holonomy of the LWY metrics is
not contained in $Sp(1)\times Sp(1)$, and so must be $Sp(2)$, whenever
$U^{\infty}$ is non-diagonal. This is sufficient to show that a metric
corresponding to two non-orthogonal 3-planes has $Sp(2)$ holonomy. Since the
holonomy cannot be reduced by the inclusion of additional 3-planes, a
solution of the \hk\ conditions that includes any two non-orthogonal 3-planes
yields a metric which also has holonomy $Sp(2)$. Thus, the only toric \hk\
8-metrics for which the holonomy is a proper subgroup of $Sp(2)$ are those
corresponding to the orthogonal intersection of two 3-planes, or two sets of
parallel 3-planes, in which case  the metric is the product of two \hk\
$4$-metrics. 

Finally, we note that for $n=2$ the three K{\" a}hler 2-forms can be 
expressed simply in terms of the three covariantly constant real chiral $SO(8)$
spinors $\eta_r$. This is most straightforward in the frame in which
these spinors are constant. If we normalize the spinors such that
$$
(\eta_r)^T \eta_r =1 \qquad (r=1,2,3)\ ,
\eqn\spina
$$
then 
$$
\Omega^r_{ab} = {1\over2}\varepsilon^{rst} (\eta_s)^T \gamma_{ab} \eta_t
\eqn\spinb
$$
where $\gamma_{ab}$ is the antisymmetrized product of pairs of $SO(8)$
Dirac matrices. For
$n>2$ the relation between the covariantly constant spinors and the triplet of
K{\" a}hler 2-forms is more involved. We refer the reader to [\wang] for
details.

\chapter{Overlapping branes from \hk\ manifolds}

We shall consider first the solution of D=11 supergravity
for which the 4-form field strength vanishes and the 11-metric is
$$
ds^2_{11} = ds^2(\bE^{2,1}) + U_{ij} d{\bf x}^i\cdot d{\bf x}^j + 
U^{ij}(d\varphi_i + A_i)(d\varphi_j + A_j)
\eqn\twoa
$$
where $U_{ij}$ is a $2\times 2$ symmetric matrix of the form \onele\
characterizing an 8-dimensional toric \hk\ manifold ${\cal M}$.  Our first task
is to determine the number of supersymmetries preserved by this solution. This 
is
essentially an application of the methods used previously in the context of KK
compactifications of D=11 supergravity (see, for example, [\DNP]).

A 32-component real spinor of the D=11 Lorentz group has the following
decomposition into representations of $Sl(2;\bR)\times SO(8)$:
$$
{\bf 32} \rightarrow ({\bf 2},{\bf 8}_s) \oplus ({\bf 2},{\bf 8}_c)\ .
\eqn\twob
$$
The two different 8-component spinors of $SO(8)$ correspond to the two possible
$SO(8)$ chiralities. The unbroken supersymmetries correspond to singlets in the
decomposition of the above $SO(8)$ representations with respect to the holonomy
group ${\cal H}$ of ${\cal M}$. For example, the D=11 vacuum corresponds to
the choice $U=U^{(\infty)}$ for which ${\cal H}$ is trivial; in this case both
8-dimensional spinor representations decompose into 8 singlets, so that all
supersymmetries are preserved. The generic holonomy group is
$Sp(2)$, for which we have the following decomposition of the $SO(8)$ spinor
representations:
$$
\eqalign{
{\bf 8}_s &\rightarrow {\bf 5} \oplus {\bf 1}\oplus {\bf 1}\oplus {\bf 1}\cr 
{\bf 8}_c &\rightarrow {\bf 4}\oplus {\bf 4}\ .}
\eqn\twoc
$$
There are now a total of 6 singlets (three $Sl(2;\bR)$ doublets) instead of 32, 
so that the D=11 supergravity solution preserves 3/16 of the supersymmetry,
unless the holonomy happens to be a proper subgroup of $Sp(2)$ in which case
the above representations must be further decomposed. For example, the ${\bf
5}$ and ${\bf 4}$ representations of $Sp(2)$ have the decomposition
$$
\eqalign{
{\bf 5} &\rightarrow ({\bf 2},{\bf 2}) \oplus ({\bf 1},{\bf 1})\cr
{\bf 4} &\rightarrow ({\bf 2},{\bf 1}) \oplus ({\bf 1},{\bf 2}) }
\eqn\twod
$$
into representations of $Sp(1)\times Sp(1)$. We see in this case that there are
two more singlets (one $Sl(2;\bR)$ doublet), from which it follows that the
solution preserves 1/4 of the supersymmetry whenever the holonomy is
$Sp(1)\times Sp(1)$.

The solution \twoa\ of D=11 supergravity has no obvious interpretation as it
stands, but we shall see how it acquires two distinct interpretations as
overlapping or intersecting 5-branes in the context of D=10 IIB supergravity.
One such solution involves only D-5-branes and will be discussed in the
following section. Here we present a IIB solution involving both Dirichlet
($R\otimes R$) and Solitonic ($NS\otimes NS$) 5-brane charges. Given that the
D=11 fields are invariant under the transformations generated by a
$U(1)$ Killing vector field $\partial/\partial y$, the D=11 supergravity action
can be reduced to the D=10 IIA supergravity action. The KK ansatz for the
bosonic fields leading to the string-frame 10-metric is
$$
\eqalign{
ds^2_{(11)} &= e^{-{2\over3}\phi(x)}dx^\mu dx^\nu g_{\mu\nu}(x) +
e^{{4\over3}\phi(x)}\big( dy+ dx^\mu C_\mu(x)\big)^2 \cr
A_{(11)} &= A(x) + B(x)\wedge dy\ ,}
\eqn\twoe
$$
where $A_{(11)}$ is the D=11 3-form potential and $x^\mu$ are the D=10
spacetime coordinates. We read off from the right hand side the bosonic fields 
of
D=10 IIA supergravity; these are the
$NS\otimes NS$ fields $(\phi, g_{\mu\nu}, B_{\mu\nu})$ and the  $R\otimes R$
fields $(C_\mu, A_{\mu\nu\rho})$. In our case we may choose $y=\varphi^2$ to
arrive at the IIA supergravity solution for which the non-vanishing fields are
$$
\eqalign{
ds^2_{10} &= \bigg({U_{11}\over \det U}\bigg)^{1\over2} \big[ds^2(\bE^{2,1}) +
U_{ij} d{\bf x}^i\cdot d{\bf x}^j \big] + 
\bigg({1\over U_{11}\det U}\bigg)^{{1\over2}} (dz + A_1)^2\cr
\phi &= {3\over4} \log U_{11} - {3\over4}\log \det U\cr
C &= A_2 - \bigg({U_{12}\over U_{11}}\bigg) \big(A_1 + dz\big)\ , }
\eqn\twof
$$
where we have set $\varphi_1=z$. Because of the triholomorphicity of
$\partial/\partial y$ the Killing spinors are all $y$-independent and therefore
survive as Killing spinors of the reduced theory\foot{A case in which
supersymmetry is broken by dimensional reduction because this condition is not
satisfied can be found in [\lpt].}.

Since $\phi$ is $z$-independent and $C$ satisfies ${\cal L}_kC=0$, where
$k=\partial/\partial z$, which is a $U(1)$ Killing vector field, we may use the
T-duality rules of [\BHO] to map \twof\ to a IIB supergravity solution. Again,
the triholomorphicity of $k$ ensures that all Killing spinors survive.
Let the D=10 spacetime coordinates be
$$
x^\mu = (x^m,z)\qquad (m=0,1,\dots,8)
\eqn\twog
$$
where $x^m$ are D=9 spacetime coordinates and $z$ is the `duality direction'
coordinate. The T-duality rules for the $NS\otimes NS$ fields, 
mapping string-frame metric to string-frame metric, are
$$
\eqalign{
d\tilde s^2 &= \big[g_{mn}- g_{zz}^{-1}(
g_{mz}g_{nz}-B_{mz}B_{nz})\big] dx^mdx^n  + 2g_{zz}^{-1}B_{zm} dzdx^m +
g_{zz}^{-1} dz^2\cr
\tilde B &= {1\over2}dx^m\wedge dx^n\big[ B_{mn} + 2 g_{zz}^{-1}
(g_{mz}B_{nz})\big] + g_{zz}^{-1} g_{zm} dz\wedge dx^m\cr
\tilde\phi &= \phi -{1\over2} \log g_{zz} }
\eqn\twoh
$$
where we indicate the transformed fields by a tilde. These rules may be read as
a map either from IIA to IIB or vice-versa.  The only T-duality
rules for the $R\otimes R$ fields that we shall need for most of this paper
are those that map from IIA to IIB, with the IIA fields restricted by
$$
B=0\qquad i_kA=0\ ,
\eqn\twoi
$$ 
where $i_k$ indicates contraction with the Killing vector field
$k=\partial/\partial z$. Given this restriction,
the T-dual IIB $R\otimes R$ fields are\foot{Our choice of field definitions
differs slightly from that of [\BHO].}
$$
\eqalign{
\ell & = C_z \cr
B' &= \big[C_m -(g_{zz})^{-1}C_z
g_{zm}\big] dx^m\wedge dz \cr
i_kD &= A \cr}
\eqn\twoj
$$
where $\ell$ is the IIB pseudoscalar, $B'$ the $R\otimes R$ two-form potential
and $D$ is the IIB 4-form potential.  Because of the self-duality of its field
strength we need specify only the components $i_k D$ of $D$.

The non-vanishing IIB fields resulting from the application of these
T-duality rules to the IIA solution \twof\ are
$$
\eqalign{
ds^2_E &= (\det U)^{3\over4} \big[(\det U)^{-1}ds^2(\bE^{2,1}) + (\det
U)^{-1}U_{ij} d{\bf x}^i\cdot d{\bf x}^j  + dz^2\big] \cr
B_{(i)} &= A_i\wedge dz \cr
\tau &= -{U_{12}\over U_{11}} + i{\sqrt{\det U}\over U_{11}} }
\eqn\twok
$$
where 
$$
\tau \equiv \ell + ie^{-\phi_B}\qquad
B_{(1)} \equiv B \qquad
B_{(2)} \equiv  B'
\eqn\twol
$$ 
and $ds^2_E$ is the {\it Einstein-frame} metric, related to the IIB string
metric by
$$
ds^2_E = e^{-{1\over2}\phi_B}ds^2_B\ .
\eqn\twom
$$

The complex scalar field $\tau$ takes values in the upper half complex
plane, on which the group $Sl(2;\bR)$ acts naturally by fractional linear
transformations of $\tau$:
$$
\tau \rightarrow {a\tau +b\over c\tau +d} ,\qquad \pmatrix{a&b\cr c&d}\equiv S
\in Sl(2;\bR)\ .
\eqn\twon
$$
The $\tau$ field equations are invariant under this action, and the invariance
extends to the full IIB supergravity field equations with $B_{(i)}$
transforming as an $Sl(2;\bR)$ doublet while the Einstein-frame metric is
$Sl(2;\bR)$ invariant; the fermion transformation properties will not be needed
here so we omit them. This invariance allows us to find new solutions as
$Sl(2;\bR)$ transforms of any given solution. To exploit this observation we
note that 
$$
{U\over\sqrt{\det U}} = {1\over {\cal I}m\, \tau}\pmatrix{ 1 & -{\cal
R}e\,\tau \cr -{\cal R}e\, \tau & |\tau|^2}
\eqn\twoo
$$
which shows that an $Sl(2;\bR)$ transformation of $\tau$ induces the linear
$Sl(2;\bR)$ transformation $U\rightarrow (S^{-1})^TUS^{-1}$. Thus, given a
solution in which
$U$ is of the form \onele\ we may find another solution of the IIB field
equations for which $U\rightarrow S^T US$ where $S\in Sl(2;\bR)$. However,
not all of these solutions will correspond to {\it non-singular} solutions of
D=11 supergravity. In fact, as we saw earlier in the context of
$4n$-dimensional \hk\ manifolds, the $Sl(2;\bR)$ transform of a complete toric
\hk\ 8-metric is not itself complete unless $S\in Sl(2;\bZ)$. Hence, only an
$Sl(2;\bZ)$ subgroup of the $Sl(2;\bR)$ symmetry group of the IIB field
equations is available for generating new solutions if we require
non-singularity in D=11, and the solutions then generated are just particular
cases of those we have already considered. 

It is known that, unlike IIB
supergravity, IIB superstring theory is not $Sl(2;\bR)$ invariant, but it is
believed that an $Sl(2;\bZ)$ subgroup survives as a symmetry of the full
non-perturbative theory, which can be viewed as a limit of a $T^2$
compactification of M-theory (see [\pktb] for a recent review). We might
therefore have made the restriction to $Sl(2;\bZ)$ {\sl ab initio} on the 
grounds that this is in any case required by M-theory. It is notable, however,
that this restriction arises independently from the requirement that our
singular IIB intersecting  brane solutions be derivable from non-singular
solutions of D=11 supergravity. This point has been noted previously 
[\dgt,\ght,\pkta] for the `basic' p-brane solutions in D=10; the principle
clearly has some validity but it is not  entirely clear why because D=11
supergravity is itself as much an effective field theory as are the D=10
supergravity theories. It seems that D=11 supergravity incorporates some of
those features of M-theory that are responsible for the resolution of
singularities.

We now have a class of solutions of IIB supergravity specified by a set
of intersecting 3-planes. As we shall now explain these solutions can be
interpreted as overlapping 5-branes. We shall start by considering the case in
which $U$ is diagonal. In the simplest of these cases the 8-metric is
the metric product of two Euclidean Taub-Nut metrics, each of which is
determined by a harmonic function with a single pointlike singularity. Let 
$H_i = [1 + (2|{\bf x}^i|)^{-1}]\,$ be the two harmonic functions;
then 
$$
U = \pmatrix{H_1({\bf x}^1) & 0\cr 0 & H_2({\bf x}^2)}
\eqn\twop
$$
and the corresponding IIB Einstein metric is 
$$
\eqalign{
ds^2_E &= (H_1H_2)^{3\over4}\big[
(H_1H_2)^{-1} ds^2(\bE^{2,1}) + H_2^{-1} d{\bf x}^1\cdot d{\bf x}^1 \cr
&\qquad + H_1^{-1}
d{\bf x}^2\cdot d{\bf x}^2  + dz^2 \big]\ .}
\eqn\twoq
$$
This is of the form generated by the `harmonic function rule' [\tseytlin,\GKT] 
for the {\it orthogonal} intersection of two 5-branes on a 2-brane.
Specifically, the singularity of $H_1$ is the position of an NS-5-brane,
delocalized along the $z$-direction, while the singularity of $H_2$ is the
position of a D-5-brane, again delocalized along the $z$ direction. The two
5-branes overlap on a 2-brane but are otherwise orthogonal.  

To see why the singularities of $H_1$ are the positions of NS-branes and those 
of $H_2$ the positions of D-branes we may begin by examining the behaviour of 
the complex scalar $\tau$ as we approach each brane while going far away from
from the others. For example, near ${\bf x}^2={\bf 0}$ but for $|{\bf
x}^1|\rightarrow \infty$ we have
$$
U\sim \pmatrix {1 & 0\cr 0& (2|{\bf x}^2|)^{-1}} 
\eqn\twoqb
$$
which shows that the IIB string coupling constant $g_s=e^{\phi_B}$ goes to
zero as $|{\bf x^2}|\rightarrow 0$. This shows that the 5-brane at
${\bf x}^2={\bf 0}$ must be one that is visible in weakly coupled string
theory, and this is true only of the D-5-brane. In support of this conclusion
we observe that $U$ is well-approximated, in the limit just considered, by the
solution of the \hk\ conditions associated with a single 3-plane with p-vector
$(0,1)$. The corresponding solution of IIB supergravity  has the feature that
only the $R\otimes R$ 2-form potential $B_{(2)}$ is non-zero. In contrast, near
${\bf x}^1={\bf 0}$, but for $|{\bf x}^2|\rightarrow
\infty$, we have
$$
U\sim \pmatrix {(2|{\bf x}^1|)^{-1} & 0\cr 0& 1} 
\eqn\twoqa
$$
which shows that $g_s\rightarrow \infty$ as $|{\bf x}^1|\rightarrow 0$, so the
5-brane at ${\bf x}^1={\bf 0}$ cannot be a D-5-brane. In fact, it must be a
NS-5-brane because the solution of the \hk\ conditions associated with a single
3-plane with p-vector $(1,0)$ has the feature that only the $NS\otimes NS$
2-form potential $B_{(1)}$ is non-zero. 

We are now in a position to interpret the general solution with 
$U^{(\infty)}=1$.
A `single 3-plane solution' of the \hk\ conditions with p-vector $(p_1,p_2)$ is
associated with a IIB superstring 5-brane with 5-brane charge vector
$(p_1,p_2)$. This follows simply from the observation that the
$Sl(2;\bZ)$ transformation that takes a D-5-brane into a 5-brane with charge
vector $(p_1,p_2)$ also takes the D-5-brane solution into the solution with
p-vector $(p_1,p_2)$. Only p-vectors with coprime $p_1$ and $p_2$ can be found
this way, but this is precisely the restriction required by non-singularity of
the \hk\ 8-metric. Thus, there is a direct correlation between the angle at
which any given 5-brane is rotated, relative to a D-5-brane, and its 5-brane
charge\foot{Evidently, this is a consequence of the requirement that the
configuration preserve at least 3/16 of the supersymmetry.}. An instructive
case to consider is the three 5-brane solution involving a D-5-brane and an
NS-5-brane, having orthogonal overlap, and one other 5-brane. As the
orientation of the third 5-brane is changed from parallel to the D-5-brane to
parallel to the NS-5-brane it changes, chameleon-like, from a D-brane to an
NS-brane. 

The interpretation of the general solution with non-diagonal $U^{(\infty)}$ is
less clear. It might seem that the correlation between the 5-brane charges and
their orientations is altered when we allow solutions for which $U^{(\infty)}$
is not the identity matrix since the inner product determining the angle between
3-planes, and hence 5-branes, is altered, whereas the the 2-form potentials
$B_{(i)}$ that determine the 5-brane charges are unchanged. However, it is
likely that the 5-brane charges depend on the expectation value of the complex
scalar $\tau$ in such a way that the correlation is maintained. For example,
the fundamental string charge on a D-string must be proportional to $\langle
\ell\rangle$. This can be seen by considering a D-string stretched between two
parallel 3-branes; since $\langle \ell\rangle$ is the vacuum angle in the
3-brane's worldvolume, the stretched D-string must appear as a dyon with
electric charge proportional to $\langle \ell\rangle$, by the Witten effect.
It follows that the D-string must carry the same fraction of fundamental string
charge, and Dirac quantization between strings and 5-branes then implies that
the NS-5-brane must carry the same fraction of D-5-brane charge. If we start
from $U^{(\infty)}=1$ and then vary $U^{(\infty)}$ at fixed string coupling
constant such that $\ell$ goes through one period then the D-5-brane
charge on a NS-5-brane must change by one unit. If we now make use of the
invariance under shifts of $\ell$ by integral numbers of its periods to return
to $\langle\ell\rangle =0$ then we also change the NS-5-brane p-vector from
$(1,0)$ to $(1,1)$. Since $U^{(\infty)}$ is again the identity matrix the
interpretation of the configuration is as given previously; that is, the angle
between the branes is given by the charge vectors. For the above process this
is equally true at all intermediate values of $U^{(\infty)}$.
This encourages us to believe that the correlation between angles and 5-brane
charge is maintained for all $U^{(\infty)}$. Given this, the interpretation of
the IIB solution \twok\ in the general case should now be clear. We have an
arbitrary number of 5-branes each specified by a 3-vector giving its distance
from the origin and a p-vector which, together with $U^{(\infty)}$, specifies
both its orientation and 5-brane charges. More generally, each 5-brane can be
replaced by a set of parallel 5-branes of the same 5-brane charge. All of these
solutions of IIB supergravity preserve 3/16 supersymmetry. 

We conclude this section by mapping the IIB solution \twok\ back to D=11 by
a different route. When this solution is T-dualized along one of the space
directions of $\bE^{2,1}$ we obtain a solution of IIA supergravity which can be
lifted back to D=11. A different, but equivalent, route to the same D=11
solution is to dimensionally reduce \twoa\ along one of the space directions of
$\bE^{2,1}$ to get the D=10 IIA supergravity solution with constant dilaton, and
metric
$$
ds^2 = ds^2(\bE^{1,1}) +  U_{ij} d{\bf x}^i\cdot d{\bf x}^j + 
U^{ij}(d\varphi_i + A_i)(d\varphi_j + A_j)\ ,
\eqn\twor
$$
all other fields vanishing. We may now T-dualize in both of the $\varphi_i$
directions to obtain a new IIA solution. Let $\varphi^i$ be the coordinates of 
the torus dual to the one with coordinates $\varphi_i$. Since all fields are
of $NS\otimes NS$ type we need only the T-duality rules of \twoh, which yield
$$
\eqalign{
ds^2 &= ds^2(\bE^{1,1}) +  U_{ij}\, dX^i\cdot dX^j\cr
B &= A_i\wedge d\varphi^i \cr
\phi &= {1\over2}\log \det U } 
\eqn\twos
$$
where
$$
X^i = \{{\bf x}^i,\varphi^i\}\qquad (i=1,2) \ .
\eqn\twot
$$
and $dX\cdot dX$ is the flat metric on $\bE^4$ (but note that $U$ is still
$T^2$ invariant so there is no dependence on the $\varphi^i$ coordinates).
This solution represents an arbitrary number of IIA NS-5-branes
intersecting on a string, generalizing previous orthogonal intersection
solutions of this type [\khuri]. We say `intersecting' here rather than
`overlapping' because in D=10 there is no separation between the branes
(although there is in D=11). We should emphasize that there is no actual string
on this `intersection'. 

This IIA solution can be lifted to the following solution
of D=11 supergravity: 
$$
\eqalign{
ds^2_{11} &= (\det U)^{2\over3}\big[(\det U)^{-1} ds^2(\bE^{1,1}) + (\det
U)^{-1} U_{ij}\, d X^i \cdot dX^j  + dy^2\big]\cr
F &= F_i\wedge d\varphi^i\wedge dy\ .}
\eqn\twou
$$
When $U$ is diagonal this reduces to
$$
\eqalign{
ds^2_{11} &= (H_1H_2)^{2\over3}\big[(H_1H_2)^{-1} ds^2(\bE^{1,1}) +
H_1^{-1}dX^{(2)}\cdot dX^{(2)} \cr
&\qquad + H_2^{-1} dX^{(1)}\cdot dX^{(1)} +
dy^2\big]\cr 
F&= F_i\wedge d\varphi^i\wedge dy \ .}
\eqn\twov
$$
This is the special case of the 1/4 supersymmetric `orthogonal M-5-branes
overlapping on a string' solution of [\GKT] for which the harmonic functions
$H_i$ are harmonic on the $i$th copy of $\bE^3$, rather than on the $i$th copy 
of
$\bE^4$. When only $\Delta U$ is diagonal, i.e. when $U^{(\infty)}$ is not,
the two fivebranes are rotated away from orthogonality and an additional 1/16 of
the supersymmetry is broken. In the more general case in which $\Delta U$ is
non-diagonal the solution can be interpreted as an arbitrary number of 5-branes
intersecting at angles determined by the associated p-vectors; these angles are
restricted only by the condition that the pairs of integers $p_i$ be coprime. An
interesting question, which we do not address here is whether these 3/16
supersymmetric solutions can be generalized to allow $U$ to depend on
all eight coordinates $\{X^{(i)},\, i=1,2\}$.

\chapter{Non-orthogonal D-branes}

Returning to the IIA solution \twos, we T-dualize in the common string
direction to find an identical solution of IIB supergravity which,
consequently, still preserves 3/16 supersymmetry. This IIB solution again
represents the overlap on a string of
$NS\otimes NS$ 5-branes but it may be mapped to a similar configuration
involving only D-5-branes by the weak-strong string coupling $Z_2\subset
Sl(2;\bZ)$ duality. In this way we deduce that 
$$
\eqalign{
ds^2_E &= (\det U)^{1\over4} \big[ ds^2(\bE^{1,1}) +  
U_{ij}\, dX^i\cdot dX^j\big] \cr
B' &= A_i\wedge d\varphi^i \cr
\tau &= i\sqrt{\det U} \ ,}
\eqn\dba
$$
is also solution of IIB supergravity preserving 3/16 supersymmetry. In the
simplest case, in which $U$ is of LWY type, this solution represents the
intersection on a string of two D-5-branes, with one rotated relative to the
other by an angle $\theta $, given by \onepc. Again, we emphasize that
by `string' we mean here only to indicate the dimensionality of the 
intersection. Since the D-5-branes have this string direction in common, the
configuration is determined by the relative orientation of two 4-planes in the
8-dimensional space spanned by both\foot{Actually two 3-planes in the
supergravity solution but this is due to the partial delocalization of the two
5-branes.}. Each 4-plane can be considered as a
quaternionic line in the quaternionic plane. A quaternionic line through the
origin is specified by a 2-vector with components $(q_1,q_2)$, where
$$
q_1 = x^2 + I x^3 + J x^4 + K x^5 \qquad
q_2 = x^6 + I x^7 + J x^8 + K x^9
\eqn\dbb
$$
and $I,J,K$ are the quaternionic imaginary units. The orientation of this line
is specified by a unit quaternionic 2-vector. The relative orientation of a
second quaternionic line through the origin is specified by an element $A$ of
$U(2;\bH)\cong Sp(2)$. The corrresponding Lie algebra is spanned by $2\times 2$
quaternionic antihermitian matrices. The diagonal antihermitian matrices
generate the $Sp(1)\cong SU(2)$ rotations about the origin within a given
4-plane. The off-diagonal quaternion contains the four angles specifying the
rotation of one 4-plane relative to another in $\bE^8$. The group element 
$A$ will commute with quaternionic conjugation only if it is generated
by an element of the Lie algebra with real off-diagonal element. In this case
$$
A= \pmatrix{\cos\theta & \sin\theta \cr 
-\sin\theta & \cos\theta}\ ,
\eqn\dbd
$$
which represents a rotation by an angle $\theta$ of the (2345) 4-plane towards
the (6789) 4-plane. The $SO(1,9)$ spinor representation of this particular
$SO(8)$ rotation is
$$
R(\theta) = \exp\big\{-{1\over2}\theta(\Gamma_{26} + \Gamma_{37}
+\Gamma_{48} +\Gamma_{59})\big\}\ .
\eqn\fourc
$$

We are now in a position to make contact with the work of Berkooz,
Douglas and Leigh [\BDL]. They considered two intersecting
Dirichlet (p+q)-branes with a common q-brane overlap. According to their
analysis, each configuration of this type is associated with an element of
$SO(2p)$ describing the rotation of one (p+q)-brane relative to the other in
the 2p-dimensional `relative transverse' space (in the terminology of [\PT]).
The identity element of $SO(2p)$ corresponds to parallel branes, which preserve
1/2 the supersymmetry. Other elements correspond to rotated branes. The
only case considered explicitly in [\BDL] was an $SU(p)$ rotation, but it was
noted that the condition for unbroken supersymmetries was analogous to the
reduced holonomy condition arising in KK compactifications. Our case
corresponds to an $Sp(2)$ rotation in $SO(8)$, We shall now verify that this
leads to the preservation of 3/16 supersymmetry. 

We recall that the covariantly constant IIB chiral spinors $\epsilon^A$
($A=1,2$) in the background spacetime of a D-5-brane in the (12345) 5-plane,
must satisfy
$$
\Gamma_{012345}\epsilon^1 = \epsilon^2\ .
\eqn\foura
$$
If the spacetime includes an additional D-5-brane that is rotated into the
(16789) 5-plane by an angle $\theta$, then $\epsilon^A$ must also satisfy 
$$
R^{-1} \Gamma_{012345} R\epsilon^1 =\epsilon^2
\eqn\fourb
$$
where $R$ is the matrix in \fourc. From the particular form of this matrix we
deduce that \fourb\ is equivalent to
$$
R(-2\theta)\Gamma_{012345}\, \epsilon^1 =\epsilon^2
\eqn\fourd
$$
which, given \foura\ and \fourc, is equivalent to
$$
\exp\big\{\theta(\Gamma_{26} + \Gamma_{37}
+\Gamma_{48} +\Gamma_{59})\big\}\, \epsilon^A = \epsilon^A \qquad (A=1,2)\ .
\eqn\fourg
$$
Thus, we have to determine the number of simultaneous solutions for two
chiral spinors $\epsilon^A$ of \foura\ and \fourg.

To proceed, we choose the following representation of the $SO(1,9)$ Dirac
matrices:
$$
\eqalign{
\Gamma_0 = \gamma_5\otimes \gamma_5 \otimes i\sigma_2 & \qquad
\Gamma_1 = \gamma_5\otimes \gamma_5 \otimes \sigma_1 \cr
\Gamma_2 = \gamma_1\otimes 1 \otimes 1 & \qquad 
\Gamma_6 = \gamma_5\otimes \gamma_1 \otimes 1\cr
\Gamma_3 = \gamma_2\otimes 1 \otimes 1& \qquad 
\Gamma_7 = \gamma_5\otimes \gamma_2 \otimes 1\cr
\Gamma_4 = \gamma_3\otimes 1 \otimes 1& \qquad 
\Gamma_8 = \gamma_5\otimes \gamma_3 \otimes 1\cr
\Gamma_5 = \gamma_4\otimes 1 \otimes 1 & \qquad 
\Gamma_9 = \gamma_5\otimes \gamma_4 \otimes 1 }
\eqn\fourh 
$$
where $\gamma_1,\dots,\gamma_4$ are the $4\times 4$ $SO(4)$ Dirac matrices
and $\gamma_5$ is their product. This representation is not real but we may
choose each $\gamma_i$ to be either real or imaginary, with $\gamma_5$ real. 
The condition \foura\ now reads
$$
[\gamma_5 \otimes 1 \otimes \sigma_3] \epsilon^1 =\epsilon^2
\eqn\fouri
$$
while the chirality condition is
$$
[\gamma_5 \otimes \gamma_5 \otimes \sigma_3] \epsilon^A = \epsilon^A
\qquad (A=1,2)\ .
\eqn\fourj
$$
Let $\epsilon_\pm^A$ be the eigenspinors of $\sigma_3$. Then
$$
\epsilon_+^A = \pmatrix{\eta_+^A\cr 0} \qquad \epsilon_-^A =
\pmatrix{0\cr\eta_-^A}\ . 
\eqn\fourk
$$
As a consequence of \fouri\ and \fourj, the 16-component spinors $\eta^A_\pm$
satisfy
$$
\eqalign{
[1\otimes \gamma_5]\eta_\pm^1 &=\eta_\pm^2\cr
[\gamma_5\otimes 1]\eta_\pm^1 &=\pm\eta_\pm^2 \ ,}
\eqn\fourl
$$
while \fourg\ is now 
$$
\bigg[\prod_{i=1}^4
\exp\{{\theta\gamma_i\gamma_5\otimes\gamma_i}\}\bigg]\, \eta_\pm^A
=\eta_\pm^A \qquad (A=1,2)\ .
\eqn\fourm
$$
This is equivalent, given \fourl, to
$$
\sin 2\theta\, (1+ L)\,\eta_+^A =0\qquad
\sin\theta\, \eta_-^A =0
\eqn\fourn
$$
where
$$
L \equiv (\gamma_1\gamma_2\otimes
\gamma_1\gamma_2 + \gamma_1\gamma_3 \otimes \gamma_1\gamma_3
+ \gamma_1\gamma_4 \otimes \gamma_1\gamma_4 )\ .
\eqn\fouro
$$
In arriving at this result we have used the fact that
$$
(L^2 \mp 2L -3)\, \eta_\pm^A =0\ ,
\eqn\fouroa
$$
which follows from \fourl. 

The conditions \fourn\ are trivially satisfied when $\theta=0,\pi$, as expected,
and yield only $\eta_-^A=0$ when $\theta=\pm\pi/2$, implying that the second
D-5-brane reduces the supersymmetry by a factor of 1/2. This is again as
expected because the second D-5-brane is orthogonal to the first one when
$\theta=\pm\pi/2$, and the constraint \fourb\ is equivalent to
$$
\Gamma_{016789}\epsilon^1=\epsilon^2
\eqn\fourp
$$
as expected of a Killing spinor in the background spacetime of a D-5-brane
in the (16789) 5-plane. For all other values of $\theta$ we deduce not only
that $\eta_-^A=0$ but also that 
$$
(1+L)\eta_+^A = 0\qquad (A=1,2)\ .
\eqn\fourq
$$
Each of the spinors $\eta_+^1$ and $\eta_+^2$ is nominally 16 component but
the conditions \fourl\ imply that each has only four independent components.
Now, we see from \fouroa\ that $L$ has eigenvalues $-1,3$ when acting on
spinors $\eta_+$. Since $L$ also has vanishing trace it can be brought to the
form
$$
L = {\rm diag} (-1,-1,-1,3)
\eqn\fourqa
$$
when acting on the 4-dimensional vector space spanned by the four independent
solutions of the conditions \fourl\ for either $\eta_+^1$ or $\eta_+^2$. Thus,
\fourq\ projects out the eigenvector of $L$ with eigenvalue $3$, leaving
only 3 of the 4 independent components of $\eta_+^1$ or $\eta_+^2$. We thus
have a total of 6 Killing spinors, which should be compared to the 32 Killing
spinors of the vacuum, i.e. the intersecting D-brane configuration preserves
3/16 supersymmetry when $\sin 2\theta\ne 0$.

\chapter{Intersecting branes from \hk\ manifolds}

We now return to the D=11 solution \twoa, and generalize it to include a
membrane, i.e. we now take as our starting point the D=11 supergravity solution
$$
\eqalign{
ds^2 &= H^{-{2\over3}} ds^2(\bE^{2,1}) + H^{1\over3} \big[
 U_{ij} d{\bf x}^i\cdot d{\bf x}^j + 
U^{ij}(d\varphi_i + A_i)(d\varphi_j + A_j) \big]\cr
F &= \pm \omega(\bE^{2,1}) \wedge dH^{-1} }
\eqn\threea
$$
where $\omega(\bE^{2,1})$ is the volume form on $\bE^{2,1}$. This is still a 
solution of D=11 supergravity provided that $H$ is a harmonic function on the
\hk\ 8-manifold. Point singularites of $H$ are naturally interpreted as the
positions of parallel membranes. For our purposes we require $H$ to be
independent of the two $\varphi$ coordinates, so singularities of $H$ will
correspond to membranes delocalized on $T^2$. Such functions satisfy
$$
U^{ij}{\bf \partial}_i\cdot {\bf \partial}_j H=0\ .
\eqn\threeb
$$
Functions of the form
$$
H= H_1({\bf x}^1) + H_2({\bf x}^2)
\eqn\onep
$$
solve this equation if the $H_i$ are harmonic on $\bE^3$, but point
singularities of $H_1$ or $H_2$ would represent membranes that are delocalized
in three more directions. We expect that there exist solutions of \threeb\
representing localized membranes (apart from the delocalization on $T^2$),
although explicit solutions may be difficult to find. We would not expect the
corresponding `generalized' membrane solutions \threea\ to be non-singular
because this is already not the case for the standard membrane solution
(corresponding to $U=1$) but it seems likely (by comparison with the $U=1$
case) that the point singularities of $H$ will be horizons that are, if not
non-singular, only mildly singular. In any case, we shall investigate the dual
versions of the generalized membrane solutions, as done in the previous section
for $H=1$. 

Our first task is again to determine the number of unbroken supersymmetries when
$H\ne1$. One finds that the $SO(1,10)$ Killing spinors $\epsilon$ have the form
$$
\epsilon = H^{-{1\over6}}\pmatrix{\zeta^1\cr \zeta^2}
\eqn\threec
$$
where $\zeta^1$ and $\zeta^2$ are two 16-component covariantly constant $SO(8)$
spinors on the \hk\ 8-manifold. In addition, $\epsilon$ satisfies
$$
\Gamma_{012}\, \epsilon =\pm \epsilon
\eqn\threeca
$$
where the sign on the right hand side is the sign of $F$ in \threea, i.e. the
sign of the membrane charge. Since the product of all eleven Dirac matrices is 
the identity matrix, this constraint implies that $\zeta^1$ and
$\zeta^2$ have a definite $SO(8)$ chirality (the same for both), depending on
the sign of the membrane charge. We can now see from \twoc\ that all
supersymmetries will be broken if the $SO(8)$ chirality projection is onto the
${\bf 8}_c$ representation whereas the fraction of unbroken supersymmetry is
unchanged by the inclusion of the membrane if the projection is onto the ${\bf
8}_s$ representation. Hence, for an appropriate choice of sign of the membrane
charge the field configuration \threea\ is again a solution of D=11
supergravity with 3/16 supersymmetry. 

Proceeding as before we can now convert this D=11 configuration into a solution
of IIB supergravity preserving the same fraction of supersymmetry.
The result is
$$
\eqalign{
ds^2_E &= (\det U)^{3\over4}H^{1\over2}
\big[H^{-1}(\det U)^{-1} ds^2(\bE^{2,1}) + (\det U)^{-1} U_{ij} d{\bf x}^i\cdot 
d{\bf x}^j + H^{-1} dz^2\big] \cr 
B_{(i)} &= A_i\wedge dy\cr
\tau &= -{U_{12}\over U_{11}} + i{\sqrt{\det U}\over U_{11}}\cr
i_k D &= \omega(\bE^{2,1})\wedge dH^{-1}}
\eqn\threed
$$
The singularities of $H$ are now to be interpreted as the
locations of parallel D-3-branes, in agreement with the `harmonic function
rule'. Otherwise, the solution has the same interpretation as before except
that it is now natural to interpret the 2-brane overlap of the 5-branes as the
intersection with the 3-branes, which are therefore `stretched' between the
5-branes (i.e. along the $z$ direction), as in the configurations considered in
[\HW].

To obtain the corresponding generalization of \twos\ we first
dimensionally reduce \threea\ to obtain the following `generalized string'
solution of IIA supergravity:
$$
\eqalign{
ds^2 &= H^{-1} ds^2(\bE^{1,1}) +  U_{ij} d{\bf x}^i\cdot d{\bf x}^j + 
U^{ij}(d\varphi_i + A_i)(d\varphi_j + A_j)\cr
B &=  \omega(\bE^{1,1})H^{-1} \cr
\phi &= -{1\over2}\log H\ . }
\eqn\threee
$$
A double dualization then yields the new IIA solution 
$$
\eqalign{
ds^2 &= H^{-1}ds^2(\bE^{1,1}) +  U_{ij}dX^i\cdot dX^j\cr
B &= A_i\wedge d\varphi^i + \omega(\bE^{1,1}) H^{-1}\cr
\phi &= {1\over2}\log \det U - {1\over2}\log H \ ,}
\eqn\threef
$$
where $\varphi^i$ are the coordinates of the dual torus. As
before, this represents the intersection of NS-5-branes on a string but the
string is now an actual IIA string, represented by the harmonic function $H$.

This IIA solution can be lifted to D=11 to give the following
generalization of \twor:
$$
\eqalign{
ds^2_{11} &= H^{1\over3}(\det U)^{2\over3}\big[H^{-1}(\det U)^{-1} 
ds^2(\bE^{1,1}) \cr
&\qquad + (\det U)^{-1} U_{ij}\, d X^i \cdot dX^j  + H^{-1}
dy^2\big]\cr 
F &= \big[F_i\wedge d\varphi^i+ \omega(\bE^{1,1})
\wedge dH^{-1}\big]\wedge dy \ .}
\eqn\threeg
$$
When $U$ is of LWY type this solution represents a set of parallel M-2-branes,
located at the singularities of $H$, intersecting two (generically
non-orthogonal) M-5-branes on a string. More generally, the M-2-branes
intersect any number of M-5-branes, at the singularities of $U$, oriented at
essentially arbitrary angles. All of these D=11 supergravity solutions preserve,
generically, 3/16 of the supersymmetry.

Returning to \threef\ we may dualize along the string direction to
obtain a IIB solution that also represents the intersection of NS-5-branes,
but for which the fundamental string is replaced by a pp wave. A
$Z_2$ strong/weak coupling duality then yields the IIB solution
$$
\eqalign{
ds^2 &= (\det U)^{1\over4}\big[ dt d\sigma + H d\sigma^2 + U_{ij}dX^i\cdot
dX^j\big] \cr
B' &= A_i \wedge d\varphi^i \cr
\tau &= i\sqrt{\det U}\ , }
\eqn\threegab
$$
which generalizes \dba, and also preserves 3/16 of the supersymmetry. It 
represents the intersection on a string of an arbitrary number of D-5-branes.
There is no actual string in the intersection but there is now a pp wave in
this direction.

The inclusion of a membrane is natural in the context of D=11 backgrounds
involving \hk\ 8-metrics because the transverse space to the membrane is
8-dimensional. In contrast, there is no similarly natural way to include a D=11
M-5-brane because its transverse space is 5-dimensional. However, there is a
natural way to incorporate M-5-branes in the special case for which the \hk\
8-metric is the product of two \hk\ 4-metrics. In fact, in this case we may
naturally incorporate two M-5-branes. Since $U$ is now assumed diagonal, let
$U_{11}=U_1({\bf x}^1)$  and $U_{22}=U_2({\bf x}^2)$, and let $H_1({\bf x}^1)$
and $H_2({\bf x}^2)$ be two harmonic functions associated with the two
M-5-branes. It will be convenient to introduce two gauge potential one-forms
$\tilde A_i$ with field strength two-forms $\tilde F_i$, having the same
relation to the harmonic functions $H_i$ as $F_i$ does to the harmonic functions
$U_i$. That is
$$
\eqalign{
dU_i &= \star F_i \cr
dH_i &= \star \tilde F_i }
\eqn\threega
$$
where $\star$ is the Hodge dual on $\bE^3$. Then
$$
\eqalign{
ds^2_{11} &= (H_1H_2)^{2\over3}\bigg[(H_1H_2)^{-1} ds^2(\bE^{1,1}) +
H_1^{-1}\big[ U_2\, d{\bf x}^2\cdot d{\bf x}^2 +
U_2^{-1}(d\varphi_2 +A_2)^2\big]\cr
&\qquad +  H_2^{-1} \big[ U_1\, d{\bf x}^1\cdot d{\bf x}^1 +
U_1^{-1}(d\varphi_1 + A_1)^2\big] + dz^2\bigg]
\cr F&= \big[\tilde F_1 \wedge (d\varphi_1 + A_1) + \tilde F_2 \wedge
(d\varphi_2 + A_2)\big]
\wedge dz }
\eqn\threeh
$$
solves the D=11 field equations. Note that the Bianchi identity for $F$ is
satisfied since
$$
dF = \sum_i \big[\tilde F_i\wedge F_i]\wedge dz \equiv 0\ ,
\eqn\threeha
$$
as a four-form on $\bE^3$ vanishes identically. When both $U_1$ and $U_2$
are constant \threeh\ reduces to the orthogonal overlap of two M-5-branes on a
string, but with the harmonic functions $H_i$ restricted to be independent of
the angular coordinates $\varphi_i$. In the general case, in which all four
harmonic functions $U_i$ and $H_i$ are non constant, this solution preserves 1/8
supersymmetry provided the relative sign of the 5-brane charges is chosen
appropriately. We shall verify this after mapping the solution into a
configuration of intersecting branes. 

We first map \threeh\ to a IIB solution. Since the restrictions \twoi\ no
longer apply we should use the full T-duality rules [\BHO], but since all
intersections are orthogonal the T-dualized solution can also be deduced
from the harmonic function rule. The non-zero fields of the resulting IIB
configuration are
$$
\eqalign{
ds^2_E &= (H_1H_2U_1U_2)^{3\over4}\bigg[ (U_1U_2H_1H_2)^{-1} ds^2(\bE^{1,1}) +
(U_2H_2)^{-1}  d{\bf x}^1\cdot d{\bf x}^1 \cr
&\qquad + (U_1H_1)^{-1} d{\bf x}^2\cdot d{\bf x}^2 + (H_1H_2)^{-1} dz^2
+ (U_1U_2)^{-1} dy^2 \bigg] \cr 
B &= A_1\wedge dz + \tilde A_2\wedge dy  \cr
B' &= A_2\wedge dz + \tilde A_1\wedge dy \cr
\tau &= i\sqrt{{H_1U_2\over H_2U_1}}\ . }
\eqn\threei
$$
This represents two NS-5-branes and two D-5-branes intersecting orthogonally
according to the following pattern
$$
\matrix{ NS: {} & 1 & 2 & 3 & 4 & 5 &{} &{} &{} &{}\cr
         NS: {} & 1 & {}&{} & {}& {}& 6 & 7 & 8 & 9\cr
          D: {} & 1 & {}&{} & {}& 5 & 6 & 7 & 8 &{}\cr
          D: {} & 1 & 2 & 3 & 4 &{} &{} &{} &{} & 9}
\eqn\threej
$$

If this IIB solution is now T-dualized along the 1-direction then the
resulting IIA configuration can be lifted to D=11 to give a new solution of
D=11 supergravity representing the orthogonal intersection of 4 M-5-branes
according to the pattern
$$
\matrix{ M: {} & 1 & 2 & 3 & 4 & 5 &{} &{} &{} &{} &{}\cr
         M: {} & 1 & {}&{} & {}& {}& 6 & 7 & 8 & 9 &{}\cr
         M: {} & {}& {}&{} & {}& 5 & 6 & 7 & 8 &{} & 10\cr
         M: {} & {}& 2 & 3 & 4 &{} &{} &{} &{} & 9 & 10}
\eqn\threek
$$
Since the intersections are all orthogonal the explicit form of the solution
is determined by the `harmonic function rule' together with the information
about which coordinates each harmonic function depends on. To fully specify
the solution we need therefore give only the latter information. This follows
directly from the explicit form of the IIB solution \threei: the harmonic
function associated with a given M-5-brane depends on the coordinates that
parameterize the 3-brane intersection of the other M-5-branes. For example, the
M-5-brane in the (12345) 5-plane is associated to a harmonic function on the
(678) 3-plane. The number of supersymmetries preserved may now be determined as
follows. The four M-5-branes lead to the following four conditions on Killing
spinors $\epsilon$:
$$
\eqalign{
\Gamma_{012345}\, \epsilon &= \epsilon \qquad 
\Gamma_{016789}\, \epsilon = \epsilon \cr
\Gamma_{05678\, 10}\, \epsilon &= \epsilon \qquad 
\Gamma_{02349\, 10}\, \epsilon = \epsilon }
\eqn\threel
$$
Any three of these conditions imply the fourth, so the solution preserves 1/8
supersymmetry. Given any three M-5-branes the signs of the 5-brane charges are
irrelevant, but the sign of the charge of the fourth M-5-brane is then
determined by the requirement that no further supersymmetries be broken; if
the other sign is chosen all supersymmetries are broken. When translated back
to the original D=11 solution \threeh\ this requirement restricts the
connection of one of the \hk\ 4-metrics to be either self-dual or anti-self
dual according to the choice made for the other one and the choice of
signs of the 5-brane charges. 

When combined with the fact that the product of
all 11 Dirac matrices is the identity matrix, the conditions imply
$$
\Gamma_{059}\epsilon=-\epsilon \qquad \Gamma_{01\, 10}\epsilon=\epsilon
\eqn\threem
$$
so it is possible to further include a pair of membranes, still preserving 1/8
supersymmetry, according to the pattern
$$
\matrix{ M: {} & 1 & 2 & 3 & 4 & 5 &{} &{} &{} &{} &{}\cr
         M: {} & 1 & {}&{} & {}& {}& 6 & 7 & 8 & 9 &{}\cr
         M: {} & {}& {}&{} & {}& 5 & 6 & 7 & 8 &{} & 10\cr
         M: {} & {}& 2 & 3 & 4 &{} &{} &{} &{} & 9 & 10 \cr
         M: {} & 1 & {}& {}& {}& {}& {}& {}& {}& {}& 10 \cr
         M: {} & {}& {}& {}& {}& 5 & {}& {}& {}& 9 & {} } 
\eqn\threen
$$
What remains to be determined is on which coordinates the harmonic
functions associated to the membranes can depend. We shall not pursue this
here. At the same level of analysis we note that if we compactify in the
10-direction then we may also include a IIA D-6-brane in the (234678) 6-plane
since it also follows from \threel\ that
$$
\Gamma_{0234678}\, \epsilon =-\epsilon
\eqn\threeo
$$
A subsequent T-dualization in the 1-direction then yields the a
configuration of IIB branes intersecting according to the pattern
$$
\matrix{ NS: {} & 1 & 2 & 3 & 4 & 5 &{} &{} &{} &{}\cr
         NS: {} & 1 & {}&{} & {}& {}& 6 & 7 & 8 & 9\cr
          D: {} & 1 & {}&{} & {}& 5 & 6 & 7 & 8 &{}\cr
          D: {} & 1 & 2 & 3 & 4 &{} &{} &{} &{} & 9\cr
          W: {} & 1 & {}&{} & {}& {}&{} &{} &{} & {}\cr
          D: {} & 1 & {}& {}&{} & 5 &{} &{} &{} & 9 \cr
          D: {} & 1 & 2 & 3 & 4 &{} &6  &7  &8  & {} }
\eqn\threep
$$
where $W$ indicates a pp wave. Remarkably, this configuration preserves 1/8 of
the supersymmetry.

\chapter{Comments}

In this paper we have constructed several new classes of overlapping and
intersecting brane solutions of D=10 and D=11 supergravity theories. We have
shown that there is a remarkable latitude allowed by supersymmetry in choosing
the relative orientations in multiple brane configurations. In the IIB case,
for example, the orthogonal overlap on a 2-brane of a set of parallel
D-5-branes with a set of parallel NS-5-branes preserves 1/4 of the
supersymmetry. This multiple brane configuration can be deformed to one in
which each of an arbitrary number of branes of mixed 5-brane charges overlap at
essentially arbitrary angles, while preserving 3/16 supersymmetry. We have
found similar configurations involving only D-5-branes, but intersecting on a
string, and other configurations of multiple intersecting D=11 fivebranes, all
with 3/16 supersymmetry. In all such cases the 3/16 supersymmetry derives from
the $Sp(2)$ holonomy of \hk\ 8-metrics. 

In the case of the IIB D-5-branes we have verified by a quaternionic extension
of the analysis of [\BDL] that a pair of intersecting D-5-branes preserves 3/16
supersymmetry if their orientations are related by a rotation in an $Sp(2)$
subgroup of $SO(8)$ commuting with multiplication by a quaternion. It was
remarked in [\BDL] that the determination of the fraction of supersymmetry
preserved is analogous to the standard holonomy argument in Kaluza-Klein (KK)
compactifications. We can now see that, at least in the $Sp(2)$ case, that this
analogy is exact because the D-brane configuration corresponds to a IIB
supergravity solution that is dual to a non-singular D=11 spacetime of $Sp(2)$
holonomy.  In view of this it would be of interest to consider other subgroups
of $SO(8)$. As pointed out in [\BDL], the holonomy analogy would lead one to
expect the existence of intersecting D-brane configurations in which one D-brane
is rotated relative to another by an $SU(4)$, $G_2$ or ${\rm Spin} (7)$
rotation matrix. If so, there presumably exist corresponding solutions of IIB
supergravity preserving 1/8, 1/8 and 1/16 of the supersymmetry, respectively.
These IIB solutions would presumably have M-theory duals, in which case one is
led to wonder whether they could be non-singular (and non-compact) D=11
spacetimes of holonomy $SU(4)$,
$G_2$ or ${\rm Spin}(7)$.

In the case in which a 3-brane intersects overlapping IIB 5-branes, the fact
that the solution preserves 3/16 of the supersymmetry implies an N=3
supersymmetry of the field theory on the 2-brane intersection. A massless D=3
field theory with N=3 supersymmetry is automatically N=4 supersymmetric,
however, so we conclude that the 2-brane intersection cannot have massless
fields, i.e. it is not free to move in any direction. This
conclusion is consistent with the conclusion reached in [\HW] that 3-branes
stretched between a D-5-brane and an NS-5-brane have no moduli. It would
nevertheless be of interest to learn more about these massive D=3 field
theories on the 2-brane intersections. It seems that the breaking of N=4 to N=3
supersymmetry is associated with gauge field Chern-Simons terms [\KL]. Such
field theories typically have solitons with interesting properties that may
have a `brane within brane' interpretation which it would  be instructive to
elucidate.

One other notable feature of some of our new solutions is that the intersection
is localized within each brane. For example, in the solution of D=11
supergravity representing the intersection of a membrane with overlapping
fivebranes the membrane may be localized in all directions other than the one
separating the fivebranes. Clearly, there is scope for more semi-localized
solutions of this form, either by further duality transformations of those
constructed here or new solutions constructed by similar methods. Consider, for
example, a special case of the `generalized' membrane solution of \threea\ in
which the 8-metric $ds^2_8$ is the product of $\bE^4$ with a Euclidean Taub-Nut
metric. Reduction on the $S^1$ isometry of the latter yields a solution of IIA
supergravity in which a membrane is localized inside a 6-brane. 

Additional possibilities arise when one considers holomorphic cycles.
Consider the case of two parallel IIA 6-branes. When one includes the 11th
dimension, the transverse space is a 4-dimensional toric \hk\ manifold with an
$S^1$-invariant holomorphic minimal 2-sphere connecting the two
6-branes. An M-5-brane wrapped around this two-cycle would appear in D=10 as the
3-brane boundary of a IIA 4-brane stretched between two 6-branes. After
T-duality this becomes a 2-brane boundary of a 3-brane stretched between two
D-5-branes. Thus, holomorphic cycles provide opportunities to
generalize the HW-type IIB solutions given here to include additional 3-branes. 

A construction was given in [\ruback] of $S^1$-invariant two real dimensional
holomorphic minimal submanifolds in general 4-dimensional toric \hk\ metrics.
The construction may be generalized to
$T^2$-invariant holomorphic 4-dimensional submanifolds in toric \hk\ 8-manifolds
as follows. We start from $\bE^6=\bE^3\times \bE^3$ with coordinates 
$({\bf x}^1, {\bf x}^2)$. A given complex structure 
$$
I_{\bf n}= n_1I + n_2 J + n_3K
\eqn\com
$$
picks out a common direction in both $\bE^3$ factors. This defines a 2-plane in
$\bE^6$. Acting with the torus $T^2$ gives a 4 real dimensional submanifold
of the \hk\ 8-manifold. Using the explicit form for the complex structures
given in section 2 one readily sees that $I_{\bf n}$ rotates the tangent vectors
to the sub-manifold into themselves, which implies that it is holomorphic with
respect to that complex structure. It then follows from Wirtinger's theorem that
it is minimal.

One may ask whether these 4 real dimensional submanifolds continue to be
holomorphic and minimal in the context of our `generalized membrane' solutions
of D=11 supergravity. The case of most interest is that in which an M-5-brane
wraps a 4-cycle that is not the product of two 2-cycles. Assuming that
the effective M-5-brane Lagrangian is the 6-volume of the worldvolume in the
induced metric, and that $\gamma$ is this induced metric when $H=1$, we deduce
from the 11-metric of \threea\ that its Dirac-Nambu-Goto (DNG) Lagrangian is 
$$
L_{DNG} = (H^{-{2\over3}}) (H^{1\over 3})^2  \sqrt {-\det \gamma}
= \sqrt {-\det \gamma}\ .
\eqn\commentb 
$$
Remarkably, the factors of $H$ cancel so the 4-dimensional submanifold
will continue to be minimal.  We therefore expect additional possibilities for
intersecting brane solutions to arise in this way. Whether they will continue
to preserve the 3/16 supersymmetric can likely be deduced by the `static
probe method' of [\tseytwo].

Finally, given that there exist so many new intersecting brane solutions of
supergravity theories it will be of great interest to see what
further insights can be obtained from M-theory or superstring theory. For
example, string theory methods could be applied to the configurations of
non-orthogonal D-5-branes preserving 3/16 supersymmetry to determine the 
spectrum of BPS states on the string intersection. One might also investigate
in string theory the nature of the singularity associated with the coincidence
of intersection points of multiple brane configurations.

\vskip 1cm

\centerline{\bf Appendix: Completeness via the \hk\ quotient construction}

In this appendix we shall sketch the \hk\ quotient construction of the
toric \hk\ $4n$-metrics and use it to demonstrate their completeness. The
discussion follows the lines of [\gibbons] and some details have been clarified
in discussions with R. Goto as part of a longer project on toric \hk\ manifolds
to be published in the future. The connection between toric \hk\ metrics and
arrangements of hyperplanes was pointed out in [\dancer].  

Suppose we are given a $4N$-dimensional toric \hk\ manifold ${\cal M}_N$ with
metric $ds^2= dy^\sigma dy^\tau g_{\sigma\tau}$ and a triholomorphic $T^N$
isometry. To obtain a new $4n$-dimensional \hk\ metric ${\cal M}_n$ ($n<N)$)
by the \hk\ quotient construction we first choose a  $T^m$ subgroup
$H$ of the triholomorphic $T^N$ group. To each generator of $T^m$ 
we may associate
a triplet of moment maps $\mu_r,\, (r=1,2,3)$ such that the 
infinitesimal action
of this generator on any of the $4N$ coordinates $y^\sigma$ is 
$$
\delta y^\sigma = t\Omega_r^{\sigma\tau} {\partial \mu_r \over\partial y^\tau}
\qquad (r=1,2,3)
\eqno (A.1)
$$
where $\Omega_r^{\sigma\tau}$ are the entries of the inverse of the $r$th
K{\"a}hler 2-form $\Omega_r$, and $t$ is an infinitesimal parameter. To obtain 
the metric on ${\cal M}_n$ we simply impose the constraints
$$
\bfmu^{(\alpha)}= -{\bf a}^{(\alpha)}\, \qquad (\alpha=1,\dots,m) \, ,
\eqno (A.2)
$$
for $m$ triplets of constants ${\bf a}^{(\alpha)}$, and project orthogonally to
the orbits of $T^m$. The $4n+m$ dimensional submanifold consisting 
of the solutions of the constraints (A.2) is called the intersection of the
level sets of the moment maps. The quotient manifold ${\cal M}_n$ will be 
complete if the torus group $T^m$ acts smoothly and without fixed points on 
this intersection. Moreover, it will be invariant under the triholomorphic 
action of the quotient group $T^n = T^N/T^m$. 

To illustrate the method we start with the following flat toric \hk\ metric  on
$\bH^{(n+m)}$: 
$$  
\eqalign{                   
ds^2 = & U^{(\infty)}_{ij}d{\bf x}^i\cdot d{\bf x}^j +
(U^{(\infty)})^{ij} d\varphi_i d\varphi_j \cr
& \qquad +
\sum_{\alpha=1}^m \bigg[ {d{\bf r}^{(\alpha)}\cdot d{\bf
r}^{(\alpha)}\over 2|{\bf r}^{(\alpha)}|}  + 2|{\bf r}^{(\alpha)}|  (d
\phi_{(\alpha)} + A_{(\alpha)})^2\bigg]\ . }
\eqno (A.3) 
$$
This metric admits the triholomorphic $T^{(m+n)}$ action ($\alpha=1,\dots,m$
$i=1,\dots,n$)
$$
\eqalign{
\phi_{(\alpha)} &\rightarrow \phi_{(\alpha)} + t_{(\alpha)}\cr
\varphi_i &\rightarrow \varphi_i +t_i - p_i^{(\alpha)} t_{(\alpha)}\, .}
\eqno (A.4)
$$
We choose the $T^m$ subgroup corresponding to $t_i=0$. 
The moment maps associated to this subgroup are
$$
{\bf r}^{(\alpha)}- \sum_{i=1}^n p_i^{(\alpha)} {\bf x}^i\, . 
\eqno (A.5)
$$
 We therefore impose the constraints
$$
{\bf r}^{(\alpha)}- \sum_{i=1}^n p_i^{(\alpha)} {\bf x}^i + 
{\bf a}^{(\alpha)} = 0. 
\eqno (A.6)
$$
for some constants ${\bf a}^{(\alpha)}$, which specify the level sets
of the moment maps. We must now check that $T^m$ acts smoothly and freely 
on the intersection of these level sets.  It is clear that the 
quantities $p_i^{(\alpha)}$ must be rational because otherwise the $T^m$ action would be ergodic. If they have common divisor then one may find an element of 
$T^m$ which does not act freely on the intersection of level sets. 
Finally if two planes coincide, or if more than two planes intersect at a 
point, then the group $T^m$ will have fixed points on the intersection
of the level sets and the quotient will again be singular. For a more detailed
mathematical discussion of completeness (albeit in a slightly 
different setting) the reader is referred to [\dancer]. 

If (A.6) is now used in (A.3) to eliminate the ${\bf r}^{(\alpha)}$ in favour 
of the ${\bf x}^i$ then we get a new toric \hk\ $4n$-metric with $U$ of the 
form \onele. The quotient space continues to admit a triholomorphic $T^n$ 
action, given by $\varphi_i \rightarrow \varphi_i + t_i$.
\vskip 1cm
{\bf Acknowledgements}:
We thank E. Bergshoeff, R. Goto and A. Tseytlin for informative discussions.
G.P. was supported by a Royal Society University Research Fellowship.

\refout
\end